\documentclass[11pt,a4paper,twoside]{article}

\usepackage{graphicx}
\usepackage{tabls}
\usepackage{amsmath}
\usepackage{amssymb}
\usepackage{bm}

\newtheorem{theorem}{Theorem}[section]

\newtheorem{prop}[theorem]{Proposition}

\newtheorem{remark}[theorem]{Remark}

\begin{document}

\title{\huge\bf  Failure of Parameter Identification Based on Adaptive Synchronization Techniques}
\author{Wei LIN$^\ddagger$ ~~ and ~~ Huan-Fei MA}
\date{\it Key Laboratory of Mathematics for Nonlinear Sciences (Fudan
University), Ministry of Education, Research Center for Nonlinear
Sciences, School of Mathematical Sciences, Fudan University,
Shanghai 200433, China.}\maketitle

$^\ddagger$ To whom correspondence should be addressed:

Dr. Wei LIN,

School of Mathematical Sciences

Fudan University

220 HanDan Rd.

Shanghai 200433

CHINA

Phone: 86-21-5566-4901

Fax: 86-21-6564-2342

E-mail: \underline{\it wlin@fudan.edu.cn}

\newpage

{\bf \large Abstract}

In the paper, several concrete examples, as well as their numerical
simulations, are given to show that parameter identification based
on the so-called adaptive synchronization techniques might be failed
if those functions with parameters pending for identification in
coupled systems are designed to be mutually linearly dependent or
approximately linearly dependent on the orbit in the synchronization
manifold.  This failure might be emergent not only when the
synchronized orbit is selected to be some sort of equilibrium or
some sort of periodic oscillation, but also when it is taken as some
type of chaotic attractor produced by driving system.  This result
implies that chaotic property of driving signal is not necessary to
achievement of parameter identification.  The mechanism inducing
such a failure, as well as the bounded property of all trajectories
generated by coupled systems, is theoretically expatiated.  New
synchronization techniques are proposed to rigorously realize the
complete synchronization and parameter identification in a class of
systems where the nonlinearity is not globally Lipschitz.  In
addition, parameter identification are discussed for systems with
time delay.

~\\

PACS: 05.45.Gg, 05.40.-a, 87.10.+e.

\newpage

\section{Introduction}

The most classical phenomenon in reference to synchronization is
owing to Huygens' observation about the synchrony of pendulum clocks
\cite{Huygens}.   Since this historical discovery, synchronization
as an omnipresent technical issue has become a focal topic of great
importance in many applications.   Moreover, the basic concept
related to chaos synchronization in coupled chaotic systems was
initially introduced by Pecora and Carrol in 1990 \cite{Pec}.  Since
their seminal paper, chaos synchronization as an interesting
research topic of great potential application has been widely
investigated and consequently applied in plenty of fields, ranging
from secure communications to pattern recognitions, from complex
network dynamics to optimization of nonlinear systems, and even from
chemical reaction to brain activity analysis \cite{Syn-A}.  In
particular, a wide varieties of synchronization approaches,
including traditional linear or nonlinear feedback coupling, impulse
coupling, invariant manifold method, adaptive design coupling
techniques, and white-noise-based coupling have been fruitfully
proposed \cite{Syn-B}-\cite{noise} and several types of
synchronization, including complete synchronization, generalized
synchronization, phase synchronization, and lag synchronization,
have been introduced in succession
\cite{complete}-\cite{generalized}.

Among all the proposed coupling approaches for realization of
complete synchronization between coupled chaotic systems with or
without time delays, the newly developed adaptive design coupling
technique has aroused a great amount of attention from many
researchers \cite{Amri}-\cite{Zhou} simply due to the reported
success in unknown parameter identification.   Their explorations
have shown that unknown parameters could be identified in several
well-known chaotic systems and even in some neural network models
with or without time delays.  In particular, consider an
$n$-dimensional nonlinear system described by
\begin{equation}\label{x system}
   \dot{\bm{x}}=\bm{F}(\bm{x},\bm{p}),
\end{equation}
where $\bm{x}=(x_1,x_2,\ldots,x_n)^T\in\mathbb{R}^n$,
$\bm{F}(x,\bm{p})=\bigg(F_1(\bm{x},\bm{p}),F_2(\bm{x},\bm{p}),\dots,F_n(\bm{x},\bm{p})\bigg)^T$,
and
\begin{equation}
   F_i(\bm{x},\bm{p})=c_i(\bm{x})+\sum^{m}_{j=1}p_{ij}f_{ij}(\bm{x}), ~~~
   i=1,2,\dots,n.
\end{equation}
Here, $c_i(\bm{x})$ and $f_{ij}(\bm{x})$ are, respectively, assumed
to be some kind of real valued functions, and
$\bm{p}=\left\{p_{ij}\right\}\in \mathcal{U}\subset \mathbb{R}^n$
are $(n\cdot m)$ parameters pending for identification, in which
$\mathcal{U}$ is some bounded set.  Given the bounded driving signal
$\bm{x}(t)$ produced by system (1), its response system is designed
through
\begin{equation}\label{y system}
 \begin{array}{l}
  \dot{\bm{y}}=\bm{F}(\bm{y},\bm{q})+\bm{\epsilon}\cdot\bm{e},\\
  \dot{\epsilon_i}=-r_i e_i^2, \quad \dot{q}_{ij}=-\delta_{ij}e_i f_{ij}(\bm{y}),\\
  i=1,2,\dots,n, \  j=1,2,\dots,m,
  \end{array}
\end{equation}
where the feedback coupling $\bm{\epsilon}\cdot\bm{e}$ is in the
form of $(\epsilon_1 e_1,\epsilon_2e_2,\dots,\epsilon_ne_n)^T$,
$e_i=(y_i-x_i)$, $\bm{q}=\left\{q_{ij}\right\}$, and both $r_i$ and
$\delta_{ij}$ are arbitrarily chosen positive constants.

A question naturally arises: ``Is it possible to accurately identify
all the $(n\cdot m)$ parameters of the chaotic system provided that
the output time series of system (\ref{x system}) are experimentally
obtained?"  When the response system is designed as (\ref{y
system}), the answer to this question, as mentioned above, is
reportedly positive.  Concretely, the complete synchronization
between the driving system (\ref{x system}) and the response system
(\ref{y system}) could be always achieved; moreover, the varying
parameters $\bm{q}$ in (\ref{y system}), initiating from arbitrary
values, will be asymptotically convergent to the correct values of
the parameters $\bm{p}$ in (\ref{x system}) as time tends towards
positive infinity.  Seemingly, their theoretical arguments are based
on a delicate design for Lyapunov function, on the well-known
Lyapunov Stability theorem, and even on the LaSalle invariance
principle.

As a matter of fact, those signals produced by these
driving-and-response systems could be completely synchronized;
nevertheless, the parameter identification might be failed if those
terms with parameters pending for identification are designed to be
mutually linearly dependent or approximately linearly dependent on
the synchronized orbit.    In the paper, not only concrete examples
with their numerical simulations will be provided to illustrate such
a failure of parameter identification, but also the mechanism
inducing this failure will be anatomized.  Also, the performed
analysis will show that {\it chaotic property of synchronized orbit
in synchronization manifold (or say, chaotic property of positive
limit set of driving signal) is not always necessary to achievement
of parameter identification}.

The rest of the paper is organized as follows.  In Section 2, three
concrete examples, as well as their numerical simulations, are
consecutively given to illustrate the possible occurrence of
parameter identification failure.  The synchronized orbits in those
examples are, respectively, selected to be some sort of equilibrium,
periodic oscillation, and chaotic attractor.  The mechanism inducing
this failure, as well as the the bounded property of all
trajectories generated by the coupled systems (\ref{x system}) and
(\ref{y system}), is theoretically expatiated in Section 3.
Furthermore, in Section 4, new synchronization techniques are
further proposed to realize complete synchronization and parameter
identification in a class of polynomial systems where the
nonlinearity is not globally Lipschitz. In Section 5, parameter
identification are discussed for systems with time delay. Finally,
the paper is closed with some concluded remarks.

\section{Examples Showing Failure of Parameter Identification}

In this section, three groups of driving-and-response systems are
concretely presented to illustrate the possible occurrence of failed
parameter identification.

First, consider the Lorenz system:
\begin{eqnarray}\label{countereg1}
\dot{x}_1&=&p_1(x_2-x_1), \nonumber\\
\dot{x}_2&=&p_2x_1-x_1x_3-x_2,\\
\dot{x}_3&=&x_1x_2-p_3x_3 \nonumber
\end{eqnarray}
as a driving system.  And the corresponding response system becomes:
\begin{eqnarray}\label{rep1}
  \dot{y}_1&=&q_1(y_2-y_1)+\epsilon_1(y_1-x_1),\nonumber\\
  \dot{y}_2&=&q_2y_1-y_1y_3-y_2+\epsilon_2(y_2-x_2),\\
  \dot{y}_3&=&y_1y_2-q_3y_3+\epsilon_3(y_3-x_3),\nonumber
\end{eqnarray}
where the updating laws of $\bm{q}=(q_1,q_2,q_3)$ and
$\bm{\epsilon}=(\epsilon_1,\epsilon_2,\epsilon_3)$ are,
respectively, taken as: $\dot{q}_1=-\delta_1(y_1-x_1)(y_2-y_1)$,
$\dot{q}_2=-\delta_2(y_2-x_2)y_2$,
$\dot{q}_3=-\delta_3(y_3-x_3)(-y_3)$, and
$\dot{\epsilon}_1=-r_1(y_1-x_1)^2$,
$\dot{\epsilon}_2=-r_2(y_2-x_2)^2$,
$\dot{\epsilon}_3=-r_3(y_3-x_3)^2$.

Particularly, when the parameters are taken as $p_1=35$,
$p_2=\frac{8}{3}$, and $p_3=28$, the complete synchronization
between systems (\ref{countereg1}) and (\ref{rep1}) could be
achieved with time evolution, which is numerically shown by
Fig.1(a).  This group of parameters, which are different from the
classical parameters inducing strange attractor of the Lorenz
system, simply make the synchronized orbit generated by system
(\ref{countereg1}) become an asymptotically stable equilibrium $E$,
as is shown by Fig.1(b). If the reported analytical results are
completely correct, it could be expected that the varying parameters
$\bm{q}$ will be eventually convergent to the accurate values of the
parameters $\bm{p}=(p_1,p_2,p_3)$.   However, this is not the case.
As depicted in Fig.2, although the numerical simulation is
consistent with the expectation for the parameters $q_{2,3}$, it is
beyond the expectation for the parameters $q_1$.  Concretely, the
value of $q_1$ does not approach but always keeps a distant from the
accurate value of $p_1$ with time evolution, which means failed
parameter identification does occur for $q_1$. Intuitively, there
must exist some mechanism inducing such differences between $q_1$
and $q_{2,3}$ when adaptive synchronization techniques are taken
into account.

Secondly, construct a driving system based on the Chen's system
through:
\begin{eqnarray}\label{countereg2}
  \dot{x}_1 &=& 30(x_2-x_1)+\mathcal{Q}(x_1,x_2,p_1,p_2),  \nonumber\\
  \dot{x}_2 &=& (28-30)x_1-x_1x_3+28x_2, \\
  \dot{x}_3 &=& x_1x_2-3x_3. \nonumber
\end{eqnarray}
where the additional term
$$
\begin{array}{lll}
     \mathcal{Q}(x_1,x_2,p_1,p_2) &=& 0.1\times
     \displaystyle \left\{p_1\frac{\big(x_1\cos(0.9026)+x_2\sin(0.9026)\big)^2}{23.44^2}\right. \\
    & & \displaystyle -p_2\left.\left[
     \frac{\big(-x_1\sin(0.9026)+x_2\cos(0.9026)\big)^2}{7.19^2}-1\right]\right\},
\end{array}
$$
and both $p_1$ and $p_2$ are parameters expected to be identified.
As a matter of fact, without the term $\mathcal{Q}$, system
(\ref{countereg2}) becomes the original Chen's system admitting an
attractive periodic orbit.  As displayed in Fig.3, the projection of
this attractive periodic orbit into the $x_1$-$x_2$ plane is
approximately looked upon as an ellipse.  Thus, when $p_1=1$ and
$p_2=-1$, the term $\mathcal{Q}$ actually is an approximate formula
of this projection in the $x_1$-$x_2$ plane.

Given a driving signal $(x_1,x_2,x_3)^T$ generated by system
(\ref{countereg2}), the corresponding response system is designed to
be in the form of
\begin{eqnarray}\label{rep2}
  \dot{y}_1 &=& 30(y_2-y_1)+\mathcal{Q}(y_1,y_2,q_1,q_2)+\epsilon_1(y_1-x_1),\nonumber\\
  \dot{y}_2 &=& (28-30)y_1-y_1y_3+28y_2+\epsilon_2(y_2-x_2), \\
  \dot{y}_3 &=& y_1y_2-3y_3+\epsilon_3(y_3-x_3).  \nonumber
\end{eqnarray}
in which, according to (\ref{y system}), the updating laws of the
two varying parameters are taken as:
$$
 \begin{array}{lll}
   \dot{q}_1  &=& \displaystyle -\delta_1(y_1-x_1)\left[\frac{
   \big(y_1\cos(0.9026)+y_2\sin(0.9026)\big)^2}{23.44^2}\right], \\
   \dot{q}_2  &=& \displaystyle
   -\delta_2(y_1-x_1)\left[-\frac{\big(-y_1\sin(0.9026)+y_2\cos(0.9026)\big)^2}{7.19^2}+1\right],
 \end{array}
$$
and the adaptive techniques of coupling strengths are, respectively,
chosen as: $\dot{\epsilon}_1=-r_1(y_1-x_1)^2$, $\dot{\epsilon}_2=
-r_2(y_2-x_2)^2$, and $\dot{\epsilon}_3=-r_3(y_3-x_3)^2$.

Contrary to the expectation, both $q_1$ and $q_2$, starting from
almost every initial values, fail to approach the real values of the
parameters $p_1=1$ and $p_2=-1$, as shown in Fig.4.  This example,
as well as the first example, shows that failure does occur for
parameter identification based on the adaptive synchronization
technique when the synchronized orbit is particularly selected to be
some type of steady dynamics, such as asymptotically stable
equilibrium and attractive periodic orbit.

Instead of the above-mentioned steady synchronized orbit, the
existing numerical results \cite{Amri}-\cite{Zhou} show that
parameter identification could be always achieved when those
synchronized orbits in the synchronization manifold are designed to
be some type of chaotic attractor in advance.   Then, a question
arises: ``Is chaotic property of synchronized orbit in
synchronization manifold necessary to achievement of parameter
identification based on the adaptive techniques?"

To find out an answer to this question, consider a 4-dimensional
model developed from the original chaotic Lorenz system as a driving
system:
\begin{equation}\label{countereg3}
 \begin{array}{lll}
   \dot{x}_1 &=& p_1(x_2-x_1),\\
   \dot{x}_2 &=& p_2x_1-x_1x_3-x_2,\\
   \dot{x}_3 &=& x_1x_2-p_3x_3+p_4x_3(1+x^3_4),\\
   \dot{x}_4 &=& ax_4+b(x_1-x_3),
 \end{array}
\end{equation}
where $p_1=10$, $p_2=28$, and $p_3=\frac{8}{3}$ are three classical
parameters for the original Lorenz system to generate chaotic
attractor, and $a=-100$, $b=0.1$, $p_4=1$.  Given these parameters,
the orbit produced by system (\ref{countereg3}) in the
synchronization manifold still exhibits chaotic property in the
phase plane, which is displayed by Fig.5(a)-(b).  This chaotic
property is further verified by calculating the largest Lyapunov
exponent of the system ($\lambda_1\approx0.54274>0$), as is shown by
Fig.5(c).

Provided with the driving signal produced by system
(\ref{countereg3}), the complete synchronization between systems
(\ref{countereg3}) and its response system could be numerically
achieved as long as the response system is designed through:
\begin{equation}\label{rep3}
 \begin{array}{lll}
   \dot{y}_1 &=& q_1(y_2-y_1)+\epsilon_1(y_1-x_1),  \\
   \dot{y}_2 &=& q_2y_1-y_1y_3-y_2+\epsilon_2(y_2-x_2), \\
   \dot{y}_3 &=& y_1y_2-q_3y_3+q_4y_3(1+y_4^3)+\epsilon_3(y_3-x_3),\\
   \dot{y}_4 &=& ay_4+b(y_1-y_3)+\epsilon_4(y_4-x_4),
 \end{array}
\end{equation}
in which the updating laws of the parameters are taken as
$\dot{q}_1=-\delta_1(y_1-x_1)(y_2-y_1)$,
$\dot{q}_2=-\delta_2(y_2-x_2)y_1$,
$\dot{q}_3=-\delta_3(y_3-x_3)(-y_3)$,
$\dot{q}_4=-\delta_4(y_3-x_3)\left[y_3(1+y_4^3)\right]$, and the
adaptive coupling strengths are taken as
$\dot{\epsilon}_1=-r_1(y_1-x_1)^2$,
$\dot{\epsilon}_2=-r_2(y_2-x_2)^2$,
$\dot{\epsilon}_3=-r_3(y_3-x_3)^2$,
$\dot{\epsilon}_4=-r_4(y_4-x_4)^2$.   In spite of the success in
complete synchronization and in parameter identification for
$q_{1,2}$, it is impossible to utilize $q_{3,4}$, initiating from
almost every points, to identify the accurate values of the
parameters $p_{3,4}$ in system (\ref{countereg3}).   All these are
shown in Fig.6. Clearly, this example implies that the answer to the
above-posed question is negative.

{\bf Remark 2.1} ~ The fourth-order Runge-Kutta scheme is used to
solve all the ordinary differential equations in our numerical
simulations.

\section{The Mechanism Inducing the Failure}

On the one hand, three concrete examples in the last section show
that some parameter identification might be failed no matter what
kind of dynamical phenomenon is displayed in the synchronization
manifold.  On the other hand, many existing numerical results always
show successful parameter identification.  In order to clarify the
mechanism inducing such a seeming paradox, we perform a more
delicate argument by adopting the LaSalle invariance principle
\cite{LaSalle} as follows. Similar to \cite{Huang}, set a Lyapunov
function candidate by
\begin{equation}\label{Vfunction}
  V(\bm{e},\bm{\epsilon},\bm{q})=
   \frac{1}{2}\sum^n_{i=1}e^2_i+\frac{1}{2}\sum^n_{i=1}\sum^m_{j=1}\frac{1}{\delta_{ij}}(q_{ij}-p_{ij})^2
  +\frac{1}{2}\sum^n_{i=1}\frac{1}{r_i}(\epsilon_i+L)^2.
\end{equation}
Then, the derivative of the function
$V(\bm{e},\bm{\epsilon},\bm{q})$ along with the coupled systems
(\ref{x system}) and (\ref{y system}) could be estimated by
$$
  \dot{V}(\bm{e},\bm{\epsilon},\bm{q})\leqslant
  (nl-L)\sum^n_{i=1}e^2_i.
$$
Here, it should be pointed out that $l$ is not the locally
Lipschitiz constant of the function $F_i(\bm{x},\bm{p})$ but the
uniformly Lipschitiz constant since the bounded property of the
trajectory $\bm{y}(t)$ generated by the newly response system
(\ref{y system}) are not confirmed but pending for confirmation yet.

Now, {\it we contend that $\bm{e}(t)$, $\bm{\epsilon}(t)$, and
$\bm{q}(t)$ are bounded for all $t\geqslant t_0$, where $t_0$ is the
initial time}.  Indeed, one of the three variables is supposed to be
unbounded on $[t_0,+\infty)$, so that
$V(\bm{e}(t),\bm{\epsilon}(t),\bm{q}(t))$ is also unbounded on
$[t_0,+\infty)$ according to (\ref{Vfunction}).   On the other hand,
$V(\bm{e}(t),\bm{\epsilon}(t),\bm{q}(t))\leqslant
V(\bm{e}(t_0),\bm{\epsilon}(t_0),\bm{q}(t_0))$ simply due to
$\dot{V}(\bm{e},\bm{\epsilon},\bm{q})\leqslant 0$ for sufficiently
large $L$.  This contradiction thus implies the bounded property of
$\bm{e}(t)$, $\bm{\epsilon}(t)$, and $\bm{q}(t)$ for all $t\geqslant
t_0$.

Therefore, in light of the LaSalle invariance principle, the
trajectory $\big(\bm{e}(t), \bm{\epsilon}(t), \bm{q}(t)\big)$
initiating from any location in the phase plane will eventually
approach the largest invariant set $\mathcal{M}$ contained in the
set
$$
  \mathcal{E}=\big\{(\bm{e},\bm{\epsilon},\bm{q})~\big|~\dot{V}(\bm{e},\bm{\epsilon},\bm{q})=0
  \big\}.
$$
Then, the main concern becomes how to make a clear description of
the invariant set $\mathcal{M}$ ever contained in the set
$\mathcal{E}$ with respect to systems (\ref{x system}) and (\ref{y
system}).   To this end, a combination of systems (\ref{x system})
and (\ref{y system}) yields
\begin{equation}\label{difference}
  \begin{array}{l}
   \displaystyle \dot{e}_i=\dot{x}_i-\dot{y}_i
     =c_i(\bm{x})-c_i(\bm{y})+\sum^m_{j=1}p_{ij}f_{ij}(\bm{x})-\sum^m_{j=1}q_{ij}f_{ij}(\bm{y})-\epsilon_i(y_i-x_i)  \\
   \displaystyle =c_i(\bm{x})-c_i(\bm{y})+\sum^m_{j=1}\big[p_{ij}f_{ij}(\bm{x})-q_{ij}f_{ij}(\bm{x})\big]
   +\sum^m_{j=1} \big[q_{ij}f_{ij}(\bm{x})- q_{ij}f_{ij}(\bm{y})\big]-\epsilon_i(y_i-x_i).
  \end{array}
\end{equation}
Also, notice that $\dot{V}(\bm{e},\bm{\epsilon},\bm{q})=0$ implies
$\bm{e}=\bm{x}-\bm{y}=\bm{0}$, $\dot{\epsilon}_i(t)\equiv 0$, and
$\dot{q}_{ij}(t)\equiv 0$.  It thus follows from (\ref{difference})
that, for every orbit
$\big(\bm{e}(t),\bm{\epsilon}(t),\bm{q}(t)\big)\in\mathcal{E}$ of
the coupled systems,
\begin{equation}\label{sum}
  \sum^m_{j=1}\big[p_{ij}-q_{ij}(t)\big]f_{ij}(\bm{x}(t))=0,
\end{equation}
where each $q_{ij}(t)$ is identical to some constant $q_{ij}^*$. And
the largest invariant set contained in $\mathcal{E}$ with respect to
systems (\ref{x system}) and (\ref{y system}) is
$$
 \mathcal{M}=\big\{(\bm{e},\bm{\epsilon},\bm{q})~\big|~\bm{e}=\bm{x}-\bm{y}=\bm{0},
 ~\epsilon_i=\epsilon_i^*, ~q_{ij}=q_{ij}^* \big\},
$$
where $\bm{x}=\bm{x}(t)$ is the synchronized orbit, or
mathematically say, the positive limit set of the driving signal.
Thus, the question becomes: ``Is each $q_{ij}^*$ surely equal to
$p_{ij}$?" From (\ref{sum}), the answer to this question is
theoretically positive provided that [LIM]: {\it for any given $i$,
$\big\{f_{ij}(\bm{x}),j=1,2,\dots,m \big\}$ are linearly independent
on the synchronized orbit $\bm{x}=\bm{x}(t)$ in the synchronization
manifold}.

For an accurate definition of linearly independent or linearly
dependent functions, refer to \cite{def}.  Also, it is valuable to
mention that two functions might be linearly independent in a domain
but linearly dependent in some subset contained in this domain.  For
example, functions $g_1(s,u)=s$ and $g_2(s,u)=u^2$ are obviously
linearly independent in $\mathbb{R}^2$ but they are linearly
dependent in a parabola-like subset $\mathcal{S}_\mu=\big\{
(s,u)\in\mathbb{R}^2~|~s=\mu u^2 \big\}\subset\mathbb{R}^2$ for some
nonzero constant $\mu$.

If hypothesis [LIM] is not satisfied, for some $i=i_0$, either there
exist two nonzero functions $f_{i_0j_1}(\bm{x})$ and
$f_{i_0j_2}(\bm{x})$ linearly dependent on the orbit $\bm{x}(t)$ in
the synchronization manifold, or simply $f_{i_0j_1}(\bm{x}(t))\equiv
0$. We focus on the former case since failure of parameter
identification could be easily illustrated in the latter case.
Accordingly, $f_{i_0j_1}(\bm{x}(t))=cf_{i_0j_2}(\bm{x}(t))$ for some
nonzero constant $c$, which at most implies that
$\big[p_{i_0j_1}-q_{i_0j_1}(t)\big]+c\big[p_{i_0j_2}-q_{i_0j_2}(t)\big]=0$.
Clearly, although $q_{i_0j_1}(t)$ and $q_{i_0j_2}(t)$ are,
respectively, identical to some constants $q_{i_0j_1}^*$ and
$q_{i_0j_2}^*$, it is not certain that $q_{i_0j_1}^*=p_{i_0j_1}$ and
$q_{i_0j_2}^*=p_{i_0j_2}$.   Actually, they are totally distinct in
most cases.   Therefore, parameter identification might always be
failed if hypothesis [LIM] is not strictly satisfied in the design
of driving-and-response systems.

Next, by virtue of the argument performed above, the reason why
parameter identification fails in the three examples given in the
previous section is expatiated as follows.

For the driving-and-response systems (\ref{countereg1}) and
(\ref{rep1}) with specific parameters, the synchronized orbit
$\bm{x}^*(t)=(x_1^*(t),x_2^*(t),x_3^*(t))^T$ in the synchronization
manifold, as shown in Fig.1, is a globally asymptotical equilibrium
$E=\bm{x}^*(t)=(6.8313,6.8313,1.6667)^T$.   Substitution of
(\ref{countereg1}) into (\ref{sum}) gives
$$
   [p_1-q_1(t)][x_2^*(t)-x_1^*(t)]=0, ~
   [p_2-q_2(t)]x_1^*(t)=0,~
   [p_3-q_3(t)]x_3^*(t)=0,
$$
where each $q_i(t)$ is identical to some constant $q_i^*$ in the
invariant set $\mathcal{M}$ ($i=1,2,3$).  According to \cite{def},
each $x_i^*(t) ~(\not=0)$ is linearly independent and
$x_2^*(t)-x_1^*(t)~(\equiv 0)$ is linearly dependent.   This implies
that $q^*_{2,3}$ is identical to $p_{2,3}$ but $q^*_1$ is not
necessarily identical to $p_1$.   Therefore, $q_1(t)$, though
obeying the updating law, will not be surely convergent to $p_1$.
This illustrates the reason why parameter identification succeeds
for $q_{2,3}$ but always fails for $q_1$ as shown in Fig.2. However,
when the synchronized orbit $\bm{x}^*(t)$ with the classical
parameters is chaotic, $x_2^*(t)-x_1^*(t)$ is nonzero and thus is
linearly independent, which satisfies hypothesis [LIM].  Hence,
$q_1(t)$ will be convergent to $p_1$ almost surely, as is shown by
many existing numerical results.  In addition, when the synchronized
orbit $\bm{x}^*(t)$ unfortunately becomes the unstable equilibrium
of the original chaotic systems, $x_2^*(t)-x_1^*(t)$ is still
identical to zero violating hypothesis [LIM].  So, $q_1(t)$ still
will not be surely convergent to the accurate value of $p_1$ in such
a case.

For the coupled systems (\ref{countereg2}) and (\ref{rep2}), the
orbit $\bm{x}^*(t)$ in the synchronization manifold, as mentioned
above, is designed to be some kind of stable periodic orbit. Its
projection into the $x_1$-$x_2$ plane, which seems like an ellipse,
could be approximately expressed by the formula
$$
  \frac{\big(x_1\cos(0.9026)+x_2\sin(0.9026)\big)^2}{23.44^2}+
  \frac{\big(-x_1\sin(0.9026)+x_2\cos(0.9026)\big)^2}{7.19^2}-1 = 0.
$$
Also, substitution of (\ref{countereg2}) into (\ref{sum}) yields
$$
 \begin{array}{l}
  \displaystyle [p_1-q_1(t)]\frac{\big(x_1(t)\cos(0.9026)+x_2(t)\sin(0.9026)\big)^2}{23.44^2}+\\
  \displaystyle
  [-p_2+q_2(t)]\left\{\frac{\big(-x_1(t)\sin(0.9026)+x_2(t)\cos(0.9026)\big)^2}{7.19^2}-1\right\}=0.
 \end{array}
$$
Thus, as long as the complete synchronization between systems
(\ref{countereg2}) and (\ref{rep2}) is achieved, the orbit
$\bm{x}(t)$, as well as $\bm{y}(t)$, will approximately approach the
stable periodic orbit $\bm{x}^*(t)$.  Then, both functions
$$
\frac{\big(x_1^*(t)\cos(0.9026)+x_2^*(t)\sin(0.9026)\big)^2}{23.44^2}
~\mbox{and}~
\frac{\big(-x_1^*(t)\sin(0.9026)+x_2^*(t)\cos(0.9026)\big)^2}{7.19^2}-1
$$
are approximately linearly dependent.  This, according to the
argument performed above, means that both $q_1$ and $q_2$ are not
suitable for parameter identification, as is verified by the
simulation results in Fig.4.

Unlike the steady dynamics exhibiting in synchronization manifold in
the previous two examples, the synchronized orbit $\bm{x}^*(t)$
generated by the driving system (\ref{countereg3}) is deliberately
designed to be chaotic in the sense of possessing positive Lyapunov
exponent.   Analogously, substitution of (\ref{countereg3}) into
(\ref{sum}) produces
$$
  [p_3-q_3(t)]x_3(t)+[p_4-q_4]x_3(t)\left\{1+[x_4(t)]^3\right\}=0.
$$
It is obvious that functions $x_3$ and $x_3(1+x_4^3)$ are linearly
independent in the whole phase plane $\mathbb{R}^4$; nevertheless,
they are approximately linearly dependent on the synchronized orbit
$\bm{x}^*(t)$ because the cubic term $[x_4^*(t)]^3$ is almost equal
to zero as time $t$ is sufficiently large (see Fig.7).  Thus, this
illustrates the reason why $q_3$ and $q_4$ initiating from a mass of
points will not be convergent to the accurate values of $p_3$ and
$p_4$, respectively, in concrete numerical simulations.

In addition, consider a case that parameter $b$ in both systems
(\ref{countereg3}) and (\ref{rep3}) is selected to be zero instead
of $0.1$.  This case could be regarded as a very special example
where parameter identification may fail in spite of existence of
chaos. In such an example,  because of $x_4^*(t)\equiv 0$, functions
$x_3$ and $x_3(1+x_4^3)$ are definitely linearly dependent on the
corresponding orbit $\bm{x}^*(t)$, which violates hypothesis [LIM].
Therefore, $q_3$ and $q_4$ can not be utilized to identify the
parameters $p_3$ and $p_4$.  In a word, chaotic property of
synchronized orbit in synchronization manifold does not always
guarantee a success in parameter identification.

\begin{remark}
{\rm In the last two examples, those functions on the orbits in the
synchronization manifold are approximately linearly dependent.
Rigorously, they are still linearly independent in a mathematical
sense, so that parameter identification might be theoretically
achieved for $q_i$ correspondingly with $p_i$. However, in real
application, discretization techniques, such as the Runge-Kutta
scheme and the Euler scheme, are always taken into account in
solving the coupled continuous differential systems. Thus, owing to
the precision limit, it is unavoidable that dynamics produced by the
discretized system may not be completely consistent with the true
dynamics generated by the original system.  It is the approximate
dependence of those functions that poses some trap of local critical
point for $q_i$ and that leads to a failure of parameter
identification in the last two examples.  Therefore, {\it not only
rigorous linear-dependence of those functions with parameters
pending for identification on the synchronized orbit but also
approximate linear-dependence on the synchronized orbit should be
always avoided whenever adaptive synchronization techniques are used
in practical parameter estimation and chaos communication.}}
\end{remark}

\section{Complete Synchronization without Globally Lipschitz Condition}

In the previous section, it is pointed out that hypothesis [LIM] is
indispensable for a successful parameter identification.   In
addition, the uniform Lipschitz condition for
$\bm{F}(\bm{x},\bm{p})$ is also important in the argument performed
above for obtaining a non-positive property of
$\dot{V}(\bm{e},\bm{\epsilon},\bm{q})$.  As a matter of fact, this
uniform condition could be loosed if the bounded property of the
response system (\ref{y system}) could be priorly estimated.
However, this prior estimation could not directly follow from the
bounded property of driving system (\ref{x system}) since dynamics
of its response system with coupling term might be completely
different from the driving system which is, though, supposed to be
bounded in advance.  Then it poses a question: ``Other than the
above coupling technique and uniform Lipschitz condition, under what
kind of coupling methods and conditions on $\bm{F}(\bm{x},\bm{p})$
can one obtain a successful parameter identification rigorously?"

To this end, it is first assumed [HPT]: {\it each
$F_i(\bm{x},\bm{p})$ is homogeneous polynomial with degree no more
than two with respect to $\bm{x}$.}  As a matter of fact, a large
quantities of nonlinear systems does not satisfy globally Lipschitz
condition but are consistent with this assumption [HPT], such as the
Lorenz system and the Chen's system.

Next, notice that
$$
\begin{array}{lll}
  2y_ky_j- 2x_kx_j &=&
  (y_k-x_k)(y_j+x_j)+(y_j-x_j)(y_k+x_k) \\
  &=& 2 e_k e_j + 2x_j e_k + 2x_k e_j
\end{array}
$$
for arbitrary $k$ and $j$.   Then,  it is easy to verify that each
$e_i\big[F_i(\bm{y},\bm{p})-F_i(\bm{x},\bm{p})\big]$ can be written
as a homogeneous polynomial with degree no more than three with
respect to $\bm{e}=\bm{y}-\bm{x}$ if assumption [HPT] holds.

Reasonably, the driving signal $\bm{x}(t)$ generated by system
(\ref{x system}) is supposed to be bounded in advance.  In order to
obtain a rigorous synchronization in the system where the
nonlinearity only satisfies assumption [HPT], we re-designed the
response system as:
\begin{equation}\label{y new}
 \begin{array}{l}
  \dot{\bm{y}}=\bm{F}(\bm{y},\bm{q})+\bm{\epsilon}\cdot\bm{e}+\bm{\omega}\cdot\bm{e}^3,\\
  \dot{\epsilon_i}=-r_i e_i^2 ,~~  \dot{\omega_i}=-s_i e_i^4, \\
  \dot{q}_{ij}=-\delta_{ij}e_i f_{ij}(\bm{y}),\\
 \end{array}
\end{equation}
where
$\bm{\omega}\cdot\bm{e}^3=\big(\omega_1e_1^3,\omega_2e_2^3,\cdots,\omega_n
e_n^3\big)^T$, each $s_i$ is arbitrarily positive constant, and
other states and parameters are the same as those defined in (\ref{y
system}).

Set a Lyapunov function candidate by
$$
  H(\bm{e},\bm{\epsilon},\bm{\omega},\bm{q})
  = \frac{1}{2}\sum^n_{i=1} e^2_i+\frac{1}{2}\sum^n_{i=1}\sum^m_{j=1}\frac{1}{\delta_{ij}}(q_{ij}-p_{ij})^2
     +\frac{1}{2}\sum^n_{i=1}\frac{1}{r_i}(\epsilon_i+M)^2+
      \frac{1}{2}\sum^n_{i=1}\frac{1}{s_i}(\omega_i+N)^2.
$$
Thus, the derivative of this function along with the coupled systems
(\ref{x system}) and (\ref{y new}) yields
$$
   \dot{H}(\bm{e},\bm{\epsilon},\bm{\omega},\bm{q})(t)
   =\sum^n_{i=1}e_i(t)\big[F_i(\bm{y}(t),\bm{p})-F_i(\bm{x}(t),\bm{p})\big]-\sum^n_{i=1} M
   e_i^2(t)
   -\sum^n_{i=1} N e_i^4(t),
$$
where both $M$ and $N$ are positive numbers.   By virtue of the
conclusion on each
$e_i\big[F_i(\bm{y},\bm{p})-F_i(\bm{x},\bm{p})\big]$ obtained above,
the elementary inequality
$$
  e_i e_j e_ k\leqslant \frac{1}{6}\sum_{l=i,j,k}\big(e_l^2+e_l^4\big),
$$
and the assumed bounded property of the driving signal $\bm{x}(t)$
and parameter set $\mathcal{U}$,  we can obtain that
$\dot{H}(\bm{e},\bm{\epsilon},\bm{\omega},\bm{q})(t)\leqslant 0$ for
sufficiently large numbers $M$ and $N$.

By using the similar argument performed in the previous section, we
can easily prove that every trajectory generated by the coupled
systems (\ref{x system}) and (\ref{y new}) is not only bounded for
all $t\geqslant t_0$ but also approaching the largest invariant set
contained in
$$
 \mathcal{E}'=\big\{(\bm{e},\bm{\epsilon},\bm{\omega},\bm{q})~\big|~\dot{H}(\bm{e},\bm{\epsilon},\bm{\omega},\bm{q})=0
\big\}
$$
with respect to these coupled systems.  More precisely, the largest
invariant set becomes
$$
 \mathcal{M}'=\big\{(\bm{e},\bm{\epsilon},\bm{\omega},\bm{q})~\big|~
 \bm{e}=\bm{0},~\epsilon_i=\epsilon_i^*, ~\omega_i=\omega_i^*, ~
  q_{ij}=q_{ij}^*
\big\},
$$
where $\epsilon_i^*$, $\omega_i^*$, and $q_{ij}^*$ are some
constants dependent on the initial values of the coupled systems.
Furthermore, to achieve an accurate parameter identification between
systems (\ref{x system}) and (\ref{y new}), hypothesis [LIM] should
be still adopted.  Then, the above performed argument could be
concluded as the following proposition.

\begin{prop}
If assumptions [LIM] and [HPT] on $\bm{F}(\bm{x},\bm{p})$ are
satisfied, the complete synchronization between driving system
(\ref{x system}) and its response system (\ref{y new}) could be
surely achieved, and the parameter identification could be
accurately realized in a mathematical sense.
\end{prop}

\begin{remark}
{\rm  As mentioned above, in numerical experiment and even in real
application, not only hypothesis [LIM] should be strictly satisfied
but also the approximate linear-dependence attributed to precision
limit should be avoided.}
\end{remark}

\begin{remark}  {\rm  Assumption [HPT] on
$\bm{F}(\bm{x},\bm{p})$ could be further generalized to some other
case where globally Lipschitz condition is not fulfilled.   For
instance, one could further consider the system where the degree of
the homogeneous polynomials is larger than two, or some of those
polynomials are non-homogeneous.  However, additional coupling terms
(e.g. $\bm{e}^{2v+1},~v=2,3,\cdots$) should be added into the
response systems in order to obtain a successful synchronization and
parameter identification in a rigorous sense.}
\end{remark}

\begin{remark}
{\rm As a matter of fact, nonlinearities in the previous three
examples are not globally Lipschitz but polynomial.  Theories and
coupling techniques (i.e. Proposition 4.1 and Remark 4.3) proposed
in this section should be utilized to deal with those systems for
obtaining a successful synchronization and parameter
identification.}
\end{remark}

\section{Parameter Identification in Systems with Time-Delay}

Time-delay, as an omnipresent phenomenon, can not be neglected in
practice.  So, in this section, complete synchronization and
parameter identification in time-delayed systems via adaptive
coupling techniques are further investigated.  For simplicity,
consider a one-dimensional driving system:
\begin{equation}\label{delay-d}
   \dot{x}(t)=a f\big(x(t)\big)+ b g\big(x(t-\tau)\big),
\end{equation}
where $\tau\geqslant 0$ is a time-delay, $a$ and $b$ are parameters
pending for identification, and functions $f$ and $g$ are assumed to
be globally Lipschitz continuous with Lipschitz constants $k_f$ and
$k_g$, respectively.  Given driving signal $x(t)$ generated by
system (\ref{delay-d}), the response system is designed as
\begin{equation}\label{delay-r}
 \begin{array}{l}
   \dot{y}(t)=\alpha(t) f\big(y(t)\big)+ \beta(t) g\big(y(t-\tau)\big) + \eta(t) e(t)+ \omega(t) e(t-\delta), \\
   \dot{\alpha}(t)= - f\big(y(t)\big) e(t),~~  \dot{\beta}(t)=-g\big(y(t-\tau)\big)e(t), \\
   \dot{\eta}(t)= -e^2(t),~~\dot{\omega}(t)= -e(t)e(t-\delta),
 \end{array}
\end{equation}
where $\delta\geqslant 0$ is a time-delay induced by coupling term,
error dynamics $e(t)=y(t)-x(t)$.  The initial conditions for coupled
system (\ref{delay-d}) and (\ref{delay-r}) are chosen as $e=\phi$,
$\alpha=A$, $\beta=B$, $\eta=E$, $\omega=W\in
\mathcal{C}\triangleq\mathbb{C}([-\max\{\tau,\delta\},0],\mathbb{R})$,
in which $\mathcal{C}$ denotes the sets of all continuous functions
from $[-\max\{\tau,\delta\},0]$ to $\mathbb{R}$.

Set a Lyapunov functional candidate by
$$
\begin{array}{lll}
  \mathcal{V}(\phi,A,B,E,W) & = &
  \displaystyle \frac{1}{2}\phi^2(0)+\frac{1}{2}[A(0)-a]^2+\frac{1}{2}[B(0)-b]^2+
  \frac{1}{2}[E(0)+L]^2 \\
  &  &
  \displaystyle +\frac{1}{2}[W(0)+M]^2 +
  \bigg\{\int_{-\tau}^{0}+\int_{-\delta}^{0}\bigg\}\phi^2(s){\rm
  d}s,
\end{array}
$$
where $L$, $M$ are some proper positive constant.    Then, the
derivative of $\mathcal{V}$ along with coupled systems
(\ref{delay-d}) and (\ref{delay-r}) could be estimated by
$$
\begin{array}{lll}
  \dot{\mathcal{V}}(\phi,A,B,E,W) &\leqslant&
  \displaystyle \left(ak_f+2-\frac{L}{2}\right)\phi^2(0)+bk_g\cdot
  |\phi(0)|\cdot|\phi(-\tau)|-\phi^2(-\tau) \\
   & & \displaystyle-\frac{L}{2}\phi^2(0)-M
  \phi(0)\phi(-\delta)-\phi^2(-\delta).
\end{array}
$$
Clearly, $\dot{\mathcal{V}}(\phi,A,B,E,W)$ becomes non-positive
provided $\displaystyle L>
\max\bigg\{\frac{M^2}{2},\frac{b^2k_g^2}{2}+2ak_f+4\bigg\}$.  By
using a similar argument performed above, we can conclude that every
trajectory
$(e_t(\phi),\alpha_t(A),\beta_t(B),\eta_t(E),\omega_t(W))$, starting
from arbitrary initial condition, is surely bounded for all
$t\geqslant -\max\{\tau,\delta\}$.

Then, according to the invariance principle for the systems with
time-delay \cite{Hale}, every trajectory, as time tends towards
positive infinity, approaches the largest invariant set
$\widetilde{\mathcal{M}}$ contained in
$$
   \widetilde{\mathcal{E}}=\left\{(\phi,A,B,E,W)\in
   \underbrace{\mathcal{C}\times\cdots\times\mathcal{C}}_{5}~\bigg|~\phi(0)=\phi(-\tau)=\phi(-\delta)=0
  \right\}
$$
with respect to coupled systems (\ref{delay-d}) and (\ref{delay-r}).
This further implies that the first component of each element in
$\widetilde{\mathcal{M}}$ is identical to zero (i.e. $\phi\equiv0$)
and the others are some constant functions (i.e. $A\equiv A^*$,
$B\equiv B^*$, $E\equiv E^*$, and $W\equiv W^*$).  The accurate
values of these constant functions rest on the initial conditions of
coupled systems (\ref{delay-d}) and (\ref{delay-r}).

Parameter identification could be achieved if both equations $A^*=a$
and $B^*=b$ are valid.  However, these equations are not always
tenable although $\phi \equiv 0$ indicates a successful complete
synchronization between systems (\ref{delay-d}) and (\ref{delay-r}).
As a matter of fact, subtraction of (\ref{delay-d}) from the first
equation in (\ref{delay-r}) yields, in $\widetilde{\mathcal{M}}$,
$$
\begin{array}{lll}
  0 &=& \alpha(t)f\big(y(t)\big)-af\big(x(t)\big)+\beta(t)g\big(y(t-\tau)\big)-bg\big(x(t-\tau)\big)
   \\
  &=& [\alpha(t)-a]f\big(x(t)\big)+[\beta(t)-b]g\big(x(t-\tau)\big)  \\
  &=& [A^*-a]f\big(x(t)\big)+[B^*-b]g\big(x(t-\tau)\big),
\end{array}
$$
which follows from $e(t)=y(t)-x(t)\equiv 0$ in
$\widetilde{\mathcal{M}}$.  Now, it is clear that when functions
$f\big(x(t)\big)$ and $g\big(x(t-\tau)\big)$, on the synchronized
orbit $x(t)$ in the synchronization manifold, are linearly
dependent, both $A^*=a$ and $B^*=b$ are not certainly tenable. More
concretely, (i) when the driving signal asymptotically tends towards
some equilibrium $x(t)\equiv x^*$ of system (\ref{delay-d}), two
constant functions $f\big(x(t)\big)\equiv f(x^*)$ and
$g\big(x(t-\tau)\big)\equiv g(x^*)$ becomes linearly dependent so
that parameter identification for $a$ and $b$ will be almost surely
failed; (ii) when the synchronized orbit $x(t)$ is periodic with
period $\tau$ and both functions $f$ and $g$ are linearly dependent
in $\mathbb{R}$, parameter identification also will be failed; (iii)
when $x(t)$ is chaotic, parameter identification will be achieved in
a mathematical sense for non-constant differential functions $f$ and
$g$, and even for $f=g$ (see an example shown in Fig.8(a) where both
$f$ and $g$ are taken as sinusoid functions). However, for case
(iii), parameter identification also might be failed in numerical
simulation or in real application.  For example, in spite of chaotic
property of $x(t)$, it is likely that $x(t)\approx x(t-\tau)$ with a
small time-delay or that fluctuation of $x(t)$ seems to be
relatively steady in a macro scale.  These extraordinary cases may
lead to approximate linear-dependence between functions
$f\big(x(t)\big)$ and $g\big(x(t-\tau)\big)$, which thus results in
failure of parameter identification in numerical simulation.  See an
illustrative example in Fig.8 (b).   In addition, $\tau=0$ could be
regarded as a special case where parameter identification is always
failed provided that functions $f$ and $g$ are linearly dependent on
$x(t)$.

In conclusion, we have the following proposition on synchronization
and parameter identification for coupled systems (\ref{delay-d}) and
(\ref{delay-r}) with time-delay.

\begin{prop}
The complete synchronization between driving system (\ref{delay-d})
and its response system (\ref{delay-r}) could be surely achieved via
adaptive coupling techniques.  Furthermore, the parameter
identification could be accurately realized in a mathematical sense
provided that $f(x(t))$ and $g(x(t-\tau))$ are linearly independent
on the synchronized orbit $x(t)$ in the synchronization manifold.
\end{prop}

\begin{remark}  {\rm  With analogous
arguments but more complicated notations, the results on the driving
system (\ref{delay-d}) could be further generalized to the case
where higher dimensional driving systems and multiple parameters
identification are taken into account. However, linear-independence
of all the functions with unknown parameters on the driving signal
is crucial to a successful parameter identification.}
\end{remark}

\section{Conclusion}

In summary, concrete examples showing possible occurrence of failed
parameter identification have been numerically given in the paper.
The mechanism inducing this failure has been further rigorously
interpreted.  It has been pointed out that chaotic property of
driving system is not always crucial to achievement of parameter
identification either in a mathematical argument or in a numerical
experiment.   Actually, it is not the chaos but the hypothesis [LIM]
that guarantees a successful parameter identification based on
adaptive synchronization techniques.  However, making good use of
chaotic property might easily lead to validity of hypothesis [LIM].
Apart from linear-dependence of functions, approximate
linear-dependence of functions with parameters pending for
identification on the synchronized orbit should be always avoided in
numerical simulation and even in real application.

Furthermore, complete synchronization via new adaptive coupling
techniques in a class of polynomial systems where nonlinearity is
not globally Lipschitz has been theoretically investigated  by
virtue of the LaSalle invariance principle.   Also it has been
rigorously verified that every trajectory generated by the coupled
systems is bounded.  By these derived theoretical results, our newly
proposed technique is convinced to be a rigorous and feasible
approach for realization of complete synchronization and parameter
identification in the Lorenz-like systems.  Besides, adaptive
coupling techniques are also imported to realize parameter
identification in systems with time-delay.  Those discussion also
shows the great importance of the condition that functions with
parameters pending for identification on the synchronized orbit
should be linearly independent.

\section{Acknowledgement}

The authors are grateful to the learned referee and Prof. Jiong Ruan
for their helpful comments and suggestions.  This research was
supported by the National Natural Foundation of China Grant
No.10501008.

\newpage

\begin{figure}[htp]
\centering
\includegraphics[scale=0.5]{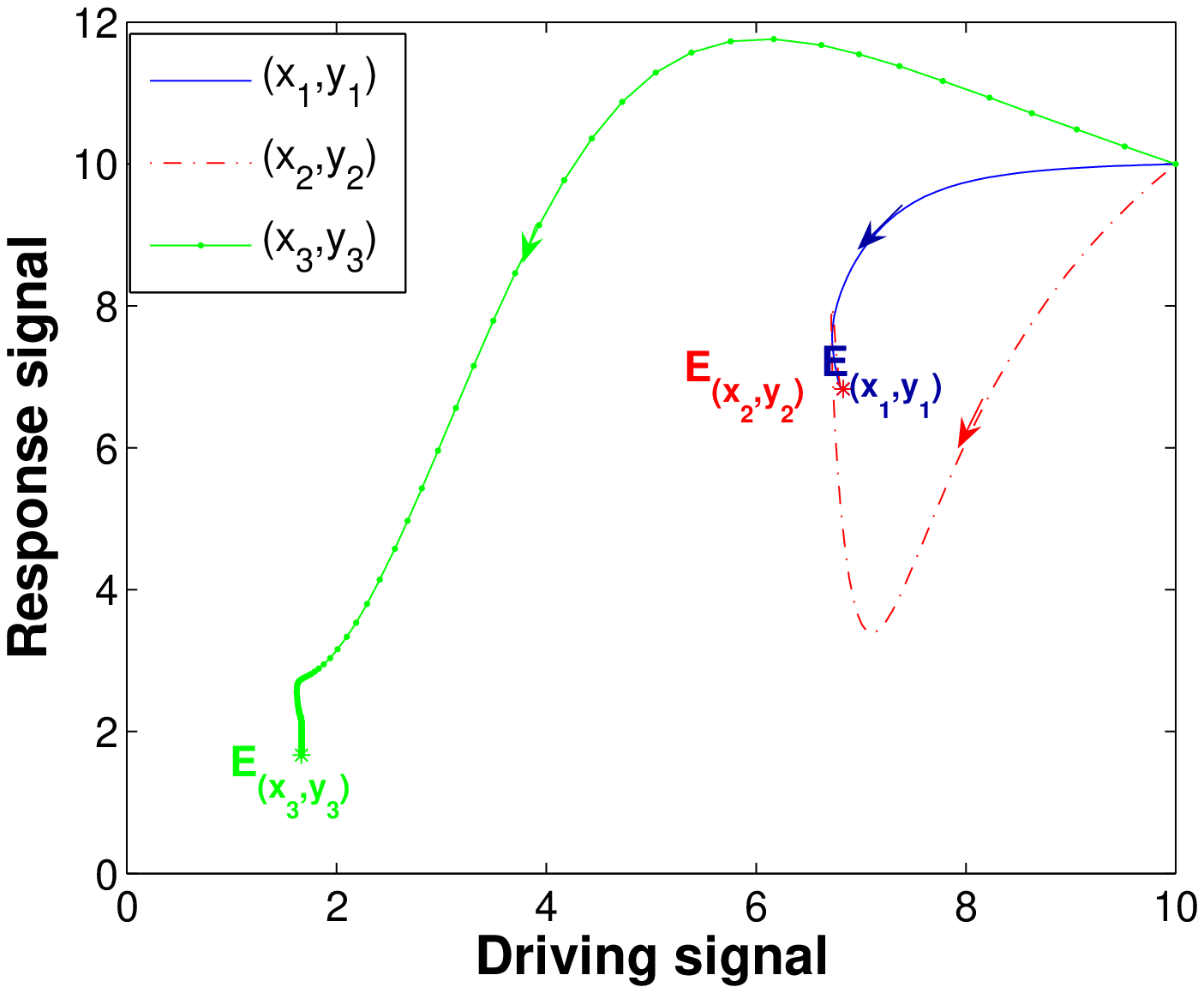}{\small(a)}
\includegraphics[scale=0.5]{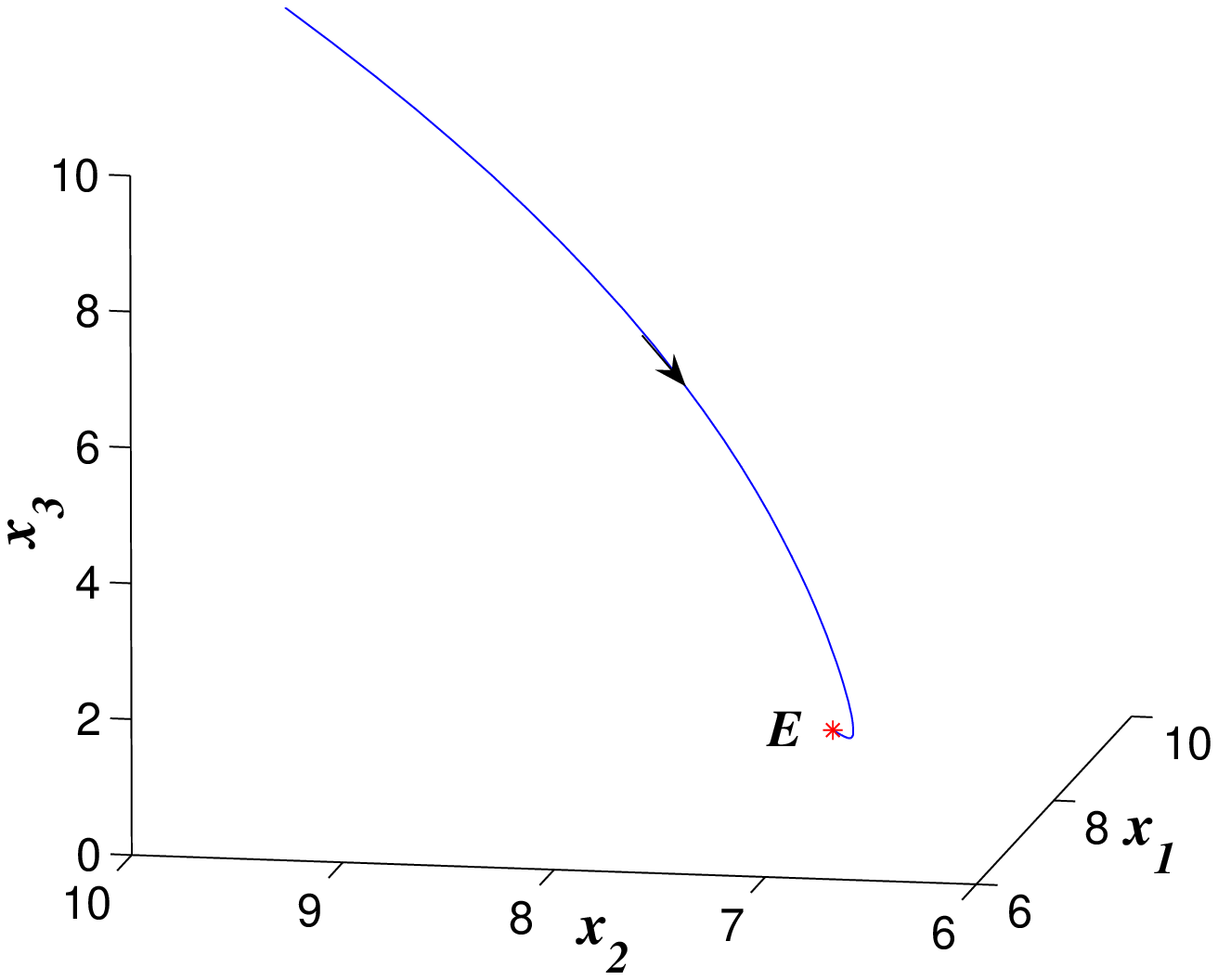}{\small(b)}
\renewcommand{\figurename}{Fig.}
\caption{\small (Color online) A successful complete synchronization
between the Lorenz system (\ref{countereg1}) and (\ref{rep1}) by
means of the adaptive design coupling. Here, system
(\ref{countereg1}) possesses an asymptotically stable equilibrium
$E=(6.8313,6.8313,1.6667)^T$ instead of the strange attractor.   The
variation of the driving signal with the response state are shown in
(a) and the evolution of response state in the phase plane are
depicted in (b). Here, $r_i=15$, $\delta_i=2$ and all the initial
values are simply chosen as $x_i^0=y_i^0=10$, $q_i^0=1$,
$\epsilon_i^0=1$ ($i=1,2,3$). }\label{fig1}
\end{figure}

\begin{figure}[htp]
\centering
\includegraphics[scale=0.5]{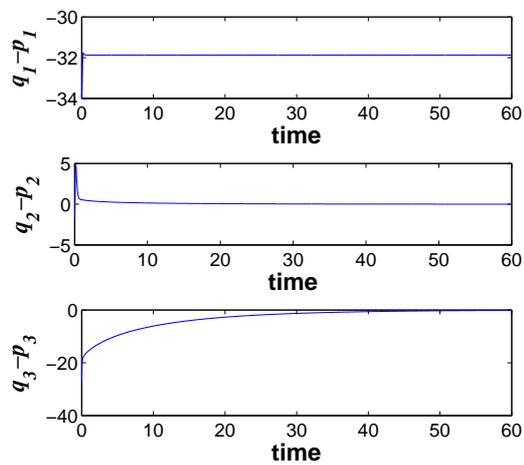}
\renewcommand{\figurename}{Fig.}
\caption{\small The variation of the error between the parameters
$q_i$ and $p_i$ with time initiating from 0 to 60 with step-size
0.01 ($i=1,2,3$).  In particular, $q_1$ fails to identify the
accurate value of $p_1$. All the parameters and initial values for
coupling systems are the same as those given in Fig.1. }\label{fig2}
\end{figure}

\begin{figure}[htp]
\centering
\includegraphics[scale=0.5]{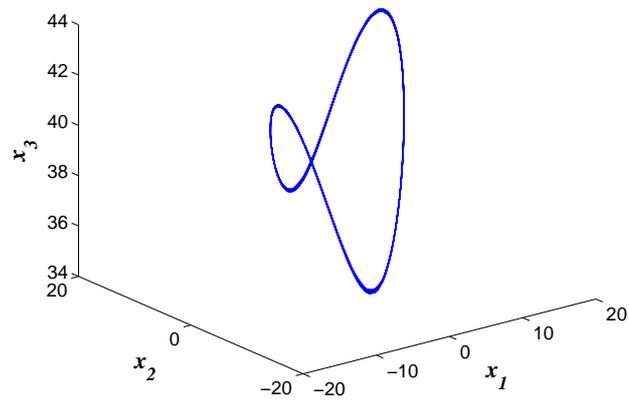} {\small(a)}
\includegraphics[scale=0.5]{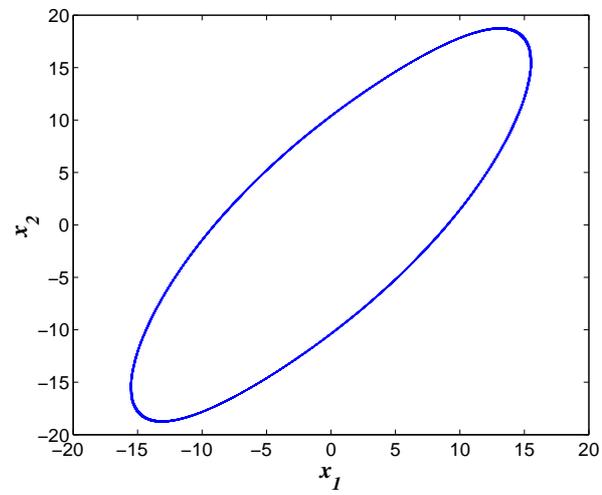} {\small(b)}
\renewcommand{\figurename}{Fig.}
\caption{\small The attractive periodic orbit generated by the
original Chen's system (system (\ref{countereg2}) when
$\mathcal{Q}\equiv 0$).  The periodic orbit in the $x_1$-$x_2$-$x_3$
phase plane (a) and its projection in the $x_1$-$x_2$ plane
(b).}\label{fig3}
\end{figure}

\begin{figure}[thp]
  \centering
  \includegraphics[scale=0.5]{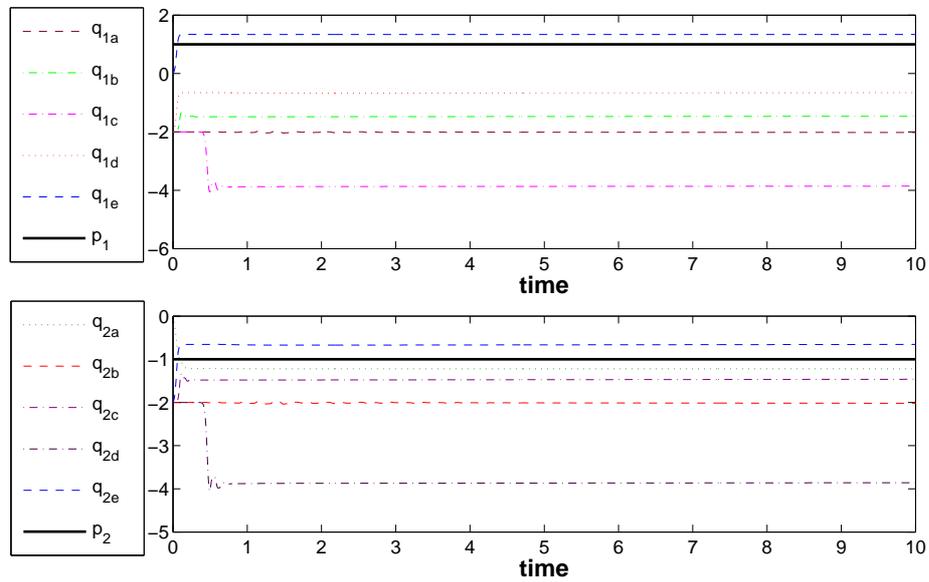}
  \renewcommand{\figurename}{Fig.}
  \caption{\small (Color
  online) The variation of $q_i$ ($i=1,2$) with time initiating from 0 to 15 with step-size
  0.01 when its initial value $q_{i\varrho}$ is differently taken ($\varrho=a,b,c,d,e$). Here, $r_j=2$, $\delta_j=1$ and all the initial
  values are taken as $x_j^0=y_j^0=10$, $\epsilon_j^0=1$, $j=1,2,3$.
  .}\label{fig4}
\end{figure}

\begin{figure}[htp]
 \centering
  \includegraphics[scale=0.5]{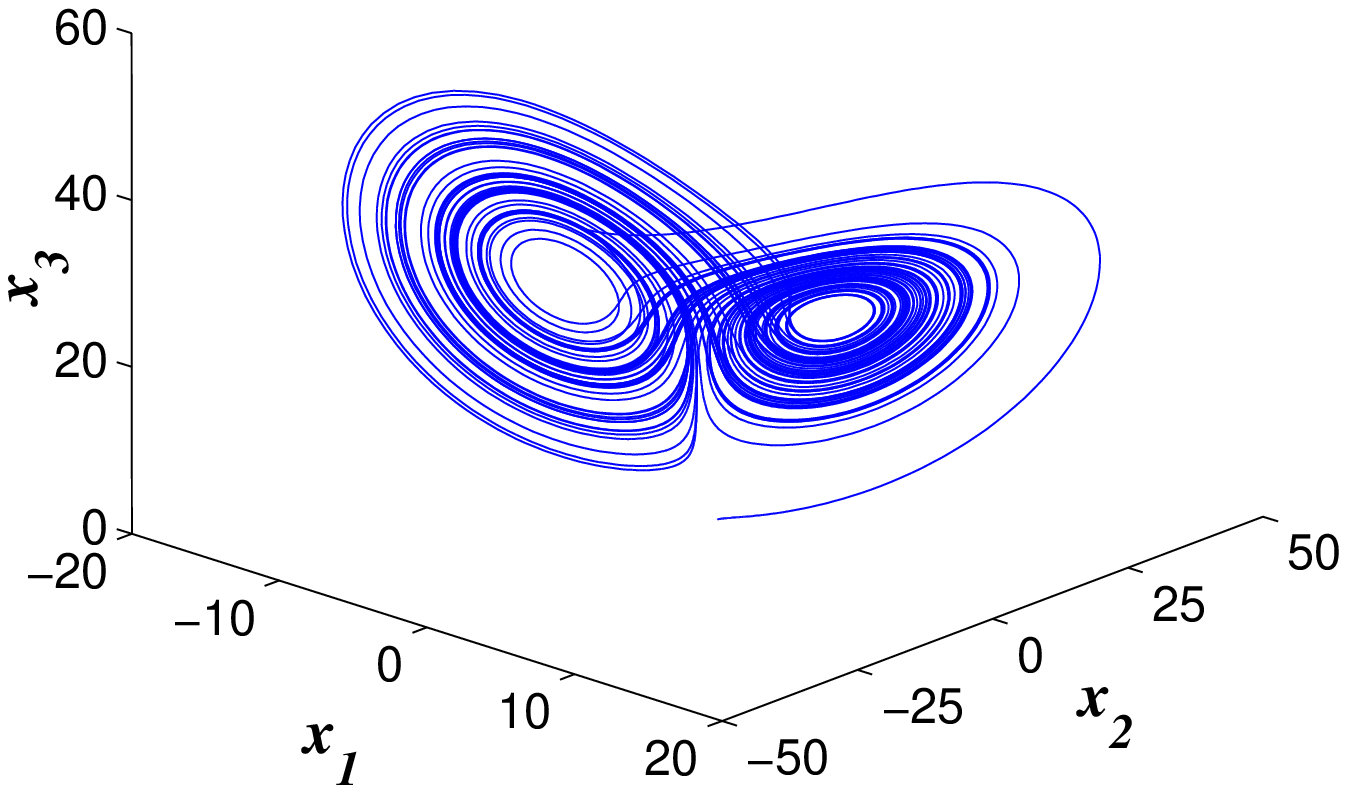}{\small (a)}
  \includegraphics[scale=0.5]{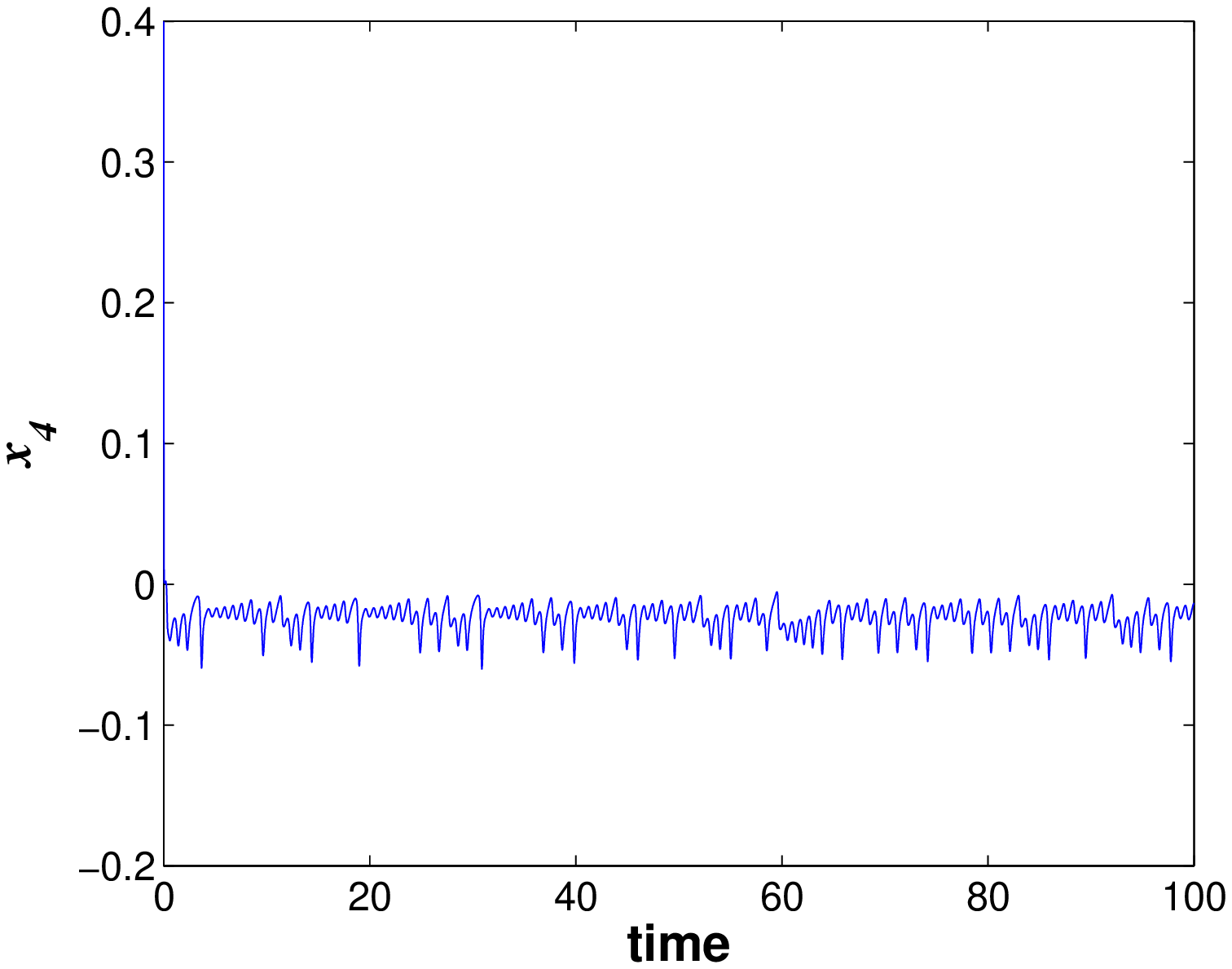}{\small (b)}
  \includegraphics[scale=0.5]{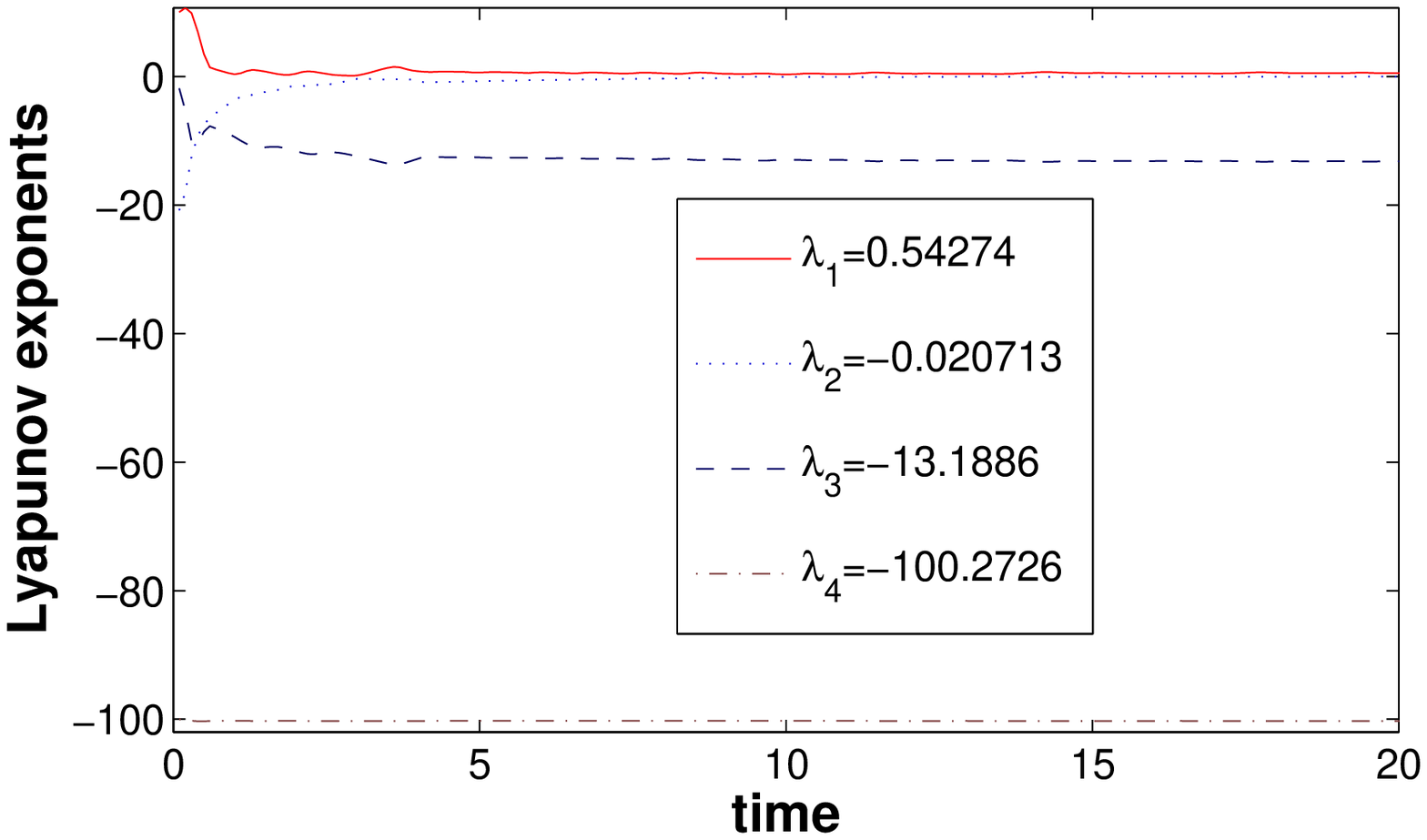}{\small (c)}
  \renewcommand{\figurename}{Fig.}
  \caption{\small (Color
  online) The strange attractor produced by system
  (\ref{countereg3}) are, respectively, plotted in the
  $x_1$-$x_2$-$x_3$ plane (a) and in the time-state-$x_4$ plane
  (b). The chaotic property is verified by the Lyapunov exponent portrait (c),
  where the largest Lyapunov exponent $\lambda_1$ is above zero.}\label{fig5}
\end{figure}

\begin{figure}[htp]
\centering
\includegraphics[scale=0.5]{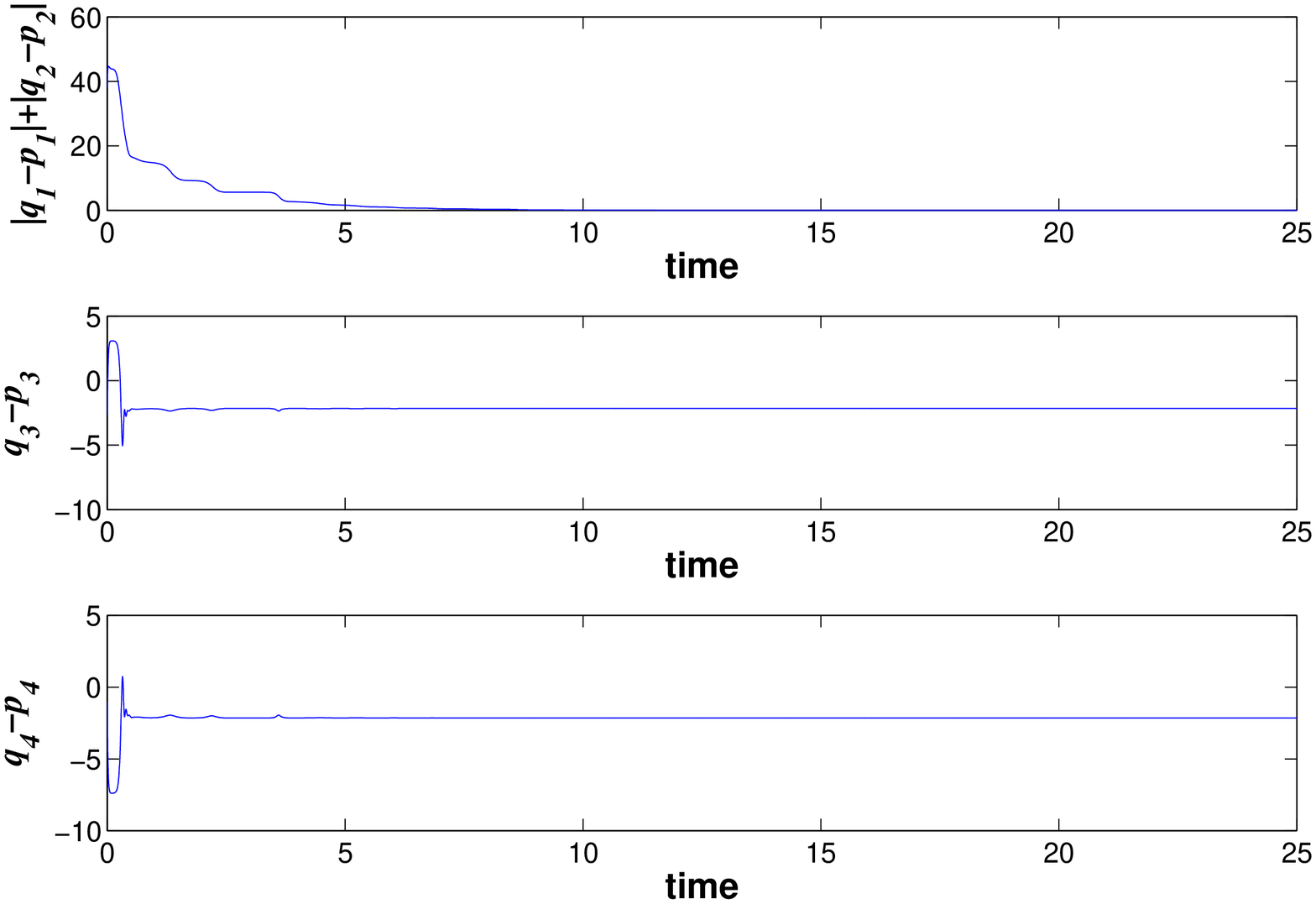}
\renewcommand{\figurename}{Fig.}
\caption{\small The variation of the error between the parameters
$q_j$ and $p_j$ with time initiating from 0 to 25 with step-size
0.01. Indeed, $q_{3,4}$ fails to identify the accurate value of
$p_{3,4}$, respectively. Here, $r_j=15$, $\delta_j=2$, and all the
initial values are taken as $x_j^0=1$, $y_1^0=6$, $y_2^0=$,
$y_3^0=10$, $y_4^0=1$, $q_j^0=0$, and $\epsilon_j^0=1$
($j=1,2,3,4$).}\label{fig6}
\end{figure}

\begin{figure}[htp]
\centering
\includegraphics[scale=0.5]{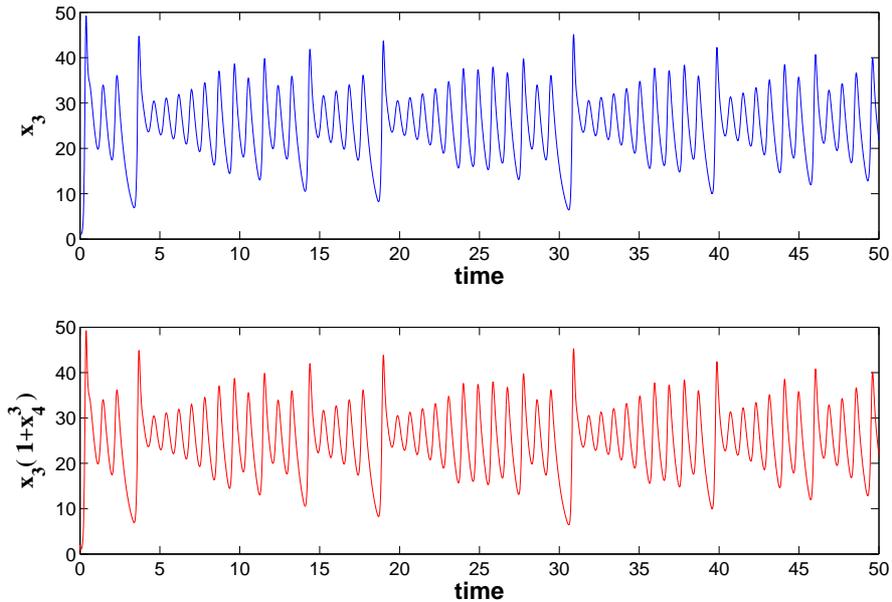}
\renewcommand{\figurename}{Fig.}  \caption{\small (Color
  online) The variation of $x_3$ and $x_3(1+x_4^3)$ on the
synchronized orbit $\bm{x}^*(t)$ with time, respectively
}\label{fig7}
\end{figure}

\begin{figure}[htp]
\centering
\includegraphics[scale=0.5]{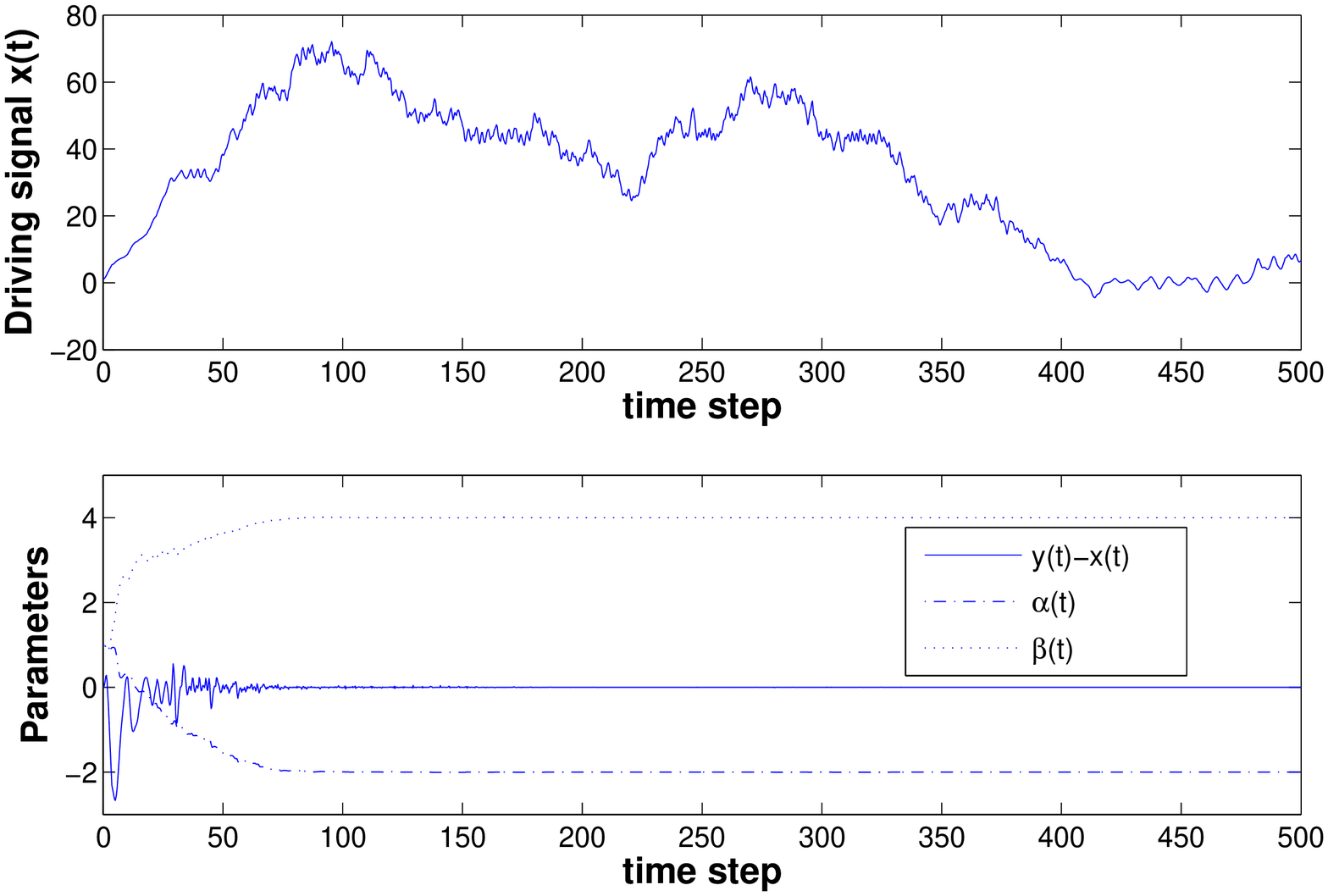}{\small (a)}
\includegraphics[scale=0.5]{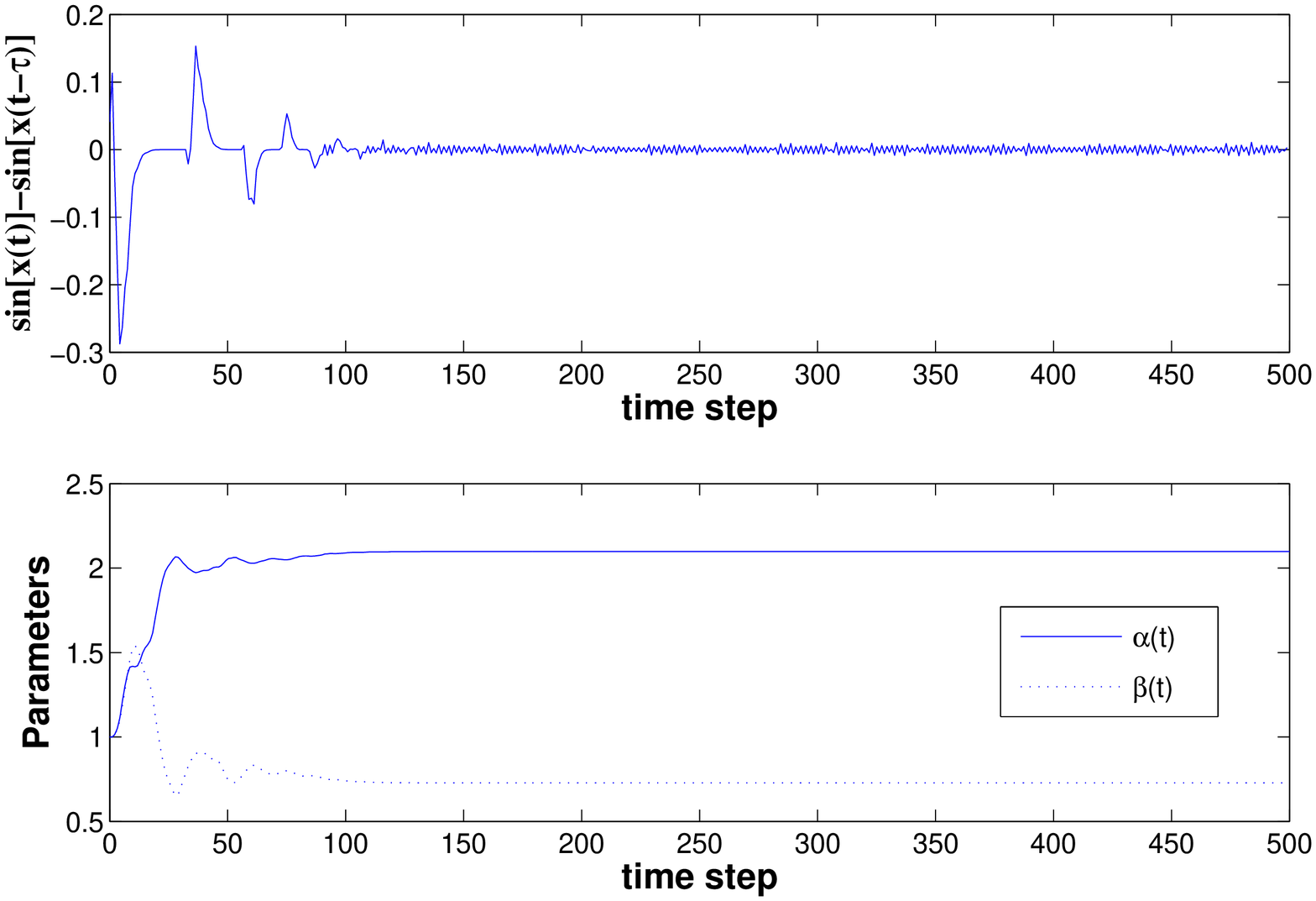}{\small (b)}
\renewcommand{\figurename}{Fig.}
\caption{\small  (a) Successful complete synchronization and
parameter identification for chaotic driving signal when $a=-2$ and
$b=4$; (b) Failed parameter identification when $a=2$ and $b=1$.
This failure is simply due to an approximate dependence between
$f\big(x(t)\big)$ and $g\big(x(t-\tau)\big)$ in a macro scale. Here,
both $f$ and $g$ are taken as sinusoid functions, time-delays are
taken as $\tau=10$, $\delta=2$, and time step size is 0.01.
}\label{fig8}
\end{figure}


\begin{thebibliography}{99}
\bibitem{def}
  A family of real valued functions $\big\{g_i(\bm{x}),i=1,2,\cdots,n\big\}$
  are said to be linearly independent in some set $\mathcal{S}$ if and
  only if $\displaystyle\sum_{i=1}^n c_i g_i(\bm{x})= 0$ (for all
  $\bm{x}\in\mathcal{S}$) implies each $c_i=0$; otherwise, they are
  said to be linearly dependent.  In particular, function
  $g(\bm{x})\equiv 0$ is linearly dependent but single nonzero
  constant function is linearly independent.
\end{thebibliography}
\end{document}